\documentclass[preprint2]{aastex6}

\received{December 30, 2016}
\revised{July 28, 2017}
\accepted{September 26, 2017}

\begin{document}


\title{Diverse nuclear star-forming activities in the heart of NGC 253 resolved with ten-pc scale ALMA images}



\author{Ryo Ando\altaffilmark{1}}
\affil{Institute of Astronomy, Graduate School of Science, The University of Tokyo, 2-21-1 Osawa, Mitaka, Tokyo 181-0015, Japan}

\author{Kouichiro Nakanishi}
\affil{National Astronomical Observatory of Japan, 2-21-1 Osawa, Mitaka, Tokyo 181-8588, Japan}
\affil{Department of Astronomy, School of Science, SOKENDAI (The Graduate University for Advanced Studies), 2-21-1 Osawa, Mitaka, Tokyo 181-8588, Japan}

\author{Kotaro Kohno}
\affil{Institute of Astronomy, Graduate School of Science, The University of Tokyo, 2-21-1 Osawa, Mitaka, Tokyo 181-0015, Japan}
\affil{Research Center for the Early Universe, Graduate School of Science, The University of Tokyo, 7-3-1 Hongo, Bunkyo, Tokyo 113-0033, Japan}

\author{Takuma Izumi}
\affil{National Astronomical Observatory of Japan, 2-21-1 Osawa, Mitaka, Tokyo 181-8588, Japan}
\affil{Institute of Astronomy, Graduate School of Science, The University of Tokyo, 2-21-1 Osawa, Mitaka, Tokyo 181-0015, Japan}

\author{Sergio Mart\'{i}n}
\affil{European Southern Observatory, Alonso de C\'{o}rdova 3107, Vitacura 763-0355, Santiago, Chile}
\affil{Joint ALMA Observatory, Alonso de C\'{o}rdova 3107, Vitacura 763-0355, Santiago, Chile}

\author{Nanase Harada}
\affil{Academia Sinica Institute of Astronomy and Astrophysics, P.O. Box 23-141, Taipei 10617, Taiwan}

\author{Shuro Takano}
\affil{Department of Physics, General Studies, College of Engineering, Nihon University,
Tamuramachi, Koriyama, Fukushima 963-8642, Japan}

\author{Nario Kuno}
\affil{Division of Physics, Faculty of Pure and Applied Sciences, University of Tsukuba, 1-1-1 Tennodai, Tsukuba, Ibaraki 305-8571, Japan}
\affil{Center for Integrated Research in Fundamental Science and Technology (CiRfSE), University of Tsukuba, Tsukuba, Ibaraki 305-8571, Japan}

\author{Naomasa Nakai}
\affil{Division of Physics, Faculty of Pure and Applied Sciences, University of Tsukuba, 1-1-1 Tennodai, Tsukuba, Ibaraki 305-8571, Japan}
\affil{Center for Integrated Research in Fundamental Science and Technology (CiRfSE), University of Tsukuba, Tsukuba, Ibaraki 305-8571, Japan}

\author{Hajime Sugai}
\affil{Kavli Institute for the Physics and Mathematics of the Universe (Kavli IPMU, WPI), The University of Tokyo, Chiba 277-8583, Japan}

\author{Kazuo Sorai}
\affil{Department of Physics, Faculty of Science, Hokkaido University, Kita 10 Nishi 8, Kita-ku, Sapporo 060-0810, Japan}
\affil{Department of Cosmosciences, Graduate School of Science, Hokkaido University, Kita 10 Nishi 8, Kita-ku, Sapporo 060-0810, Japan}

\author{Tomoka Tosaki}
\affil{Joetsu University of Education, Yamayashiki-machi, Joetsu, Niigata 943-8512, Japan}

\author{Kazuya Matsubayashi}
\affil{Okayama Astrophysical Observatory, National Astronomical Observatory of Japan, 3037-5 Honjo, Kamogata, Asakuchi, Okayama 719-0232, Japan}

\author{Taku Nakajima}
\affil{Institute for Space-Earth Environmental Research, Nagoya University, Furo-cho, Chikusa-ku, Nagoya, Aichi 464-8601, Japan}

\author{Yuri Nishimura}
\affil{Institute of Astronomy, Graduate School of Science, The University of Tokyo, 2-21-1 Osawa, Mitaka, Tokyo 181-0015, Japan}
\affil{National Astronomical Observatory of Japan, 2-21-1 Osawa, Mitaka, Tokyo 181-8588, Japan}

\and

\author{Yoichi Tamura}
\affil{Division of Particle and Astrophysical Science, Graduate School of Science, Nagoya University, Furo-cho, Chikusa-ku, Nagoya, Aichi 464-8602, Japan}
\affil{Institute of Astronomy, Graduate School of Science, The University of Tokyo, 2-21-1 Osawa, Mitaka, Tokyo 181-0015, Japan}


\altaffiltext{1}{ando@ioa.s.u-tokyo.ac.jp}

\begin{abstract}
We present an 8 pc $\times$ 5 pc resolution view of the central $\sim 200$ pc region of the nearby starburst galaxy NGC 253, based on ALMA Band 7 ($\lambda \simeq 0.85$ mm or $\nu \sim 350$ GHz) observations covering 11 GHz. We resolve the nuclear starburst of NGC 253 into eight dusty star-forming clumps, 10 pc in scale, for the first time.
These clumps, each of which contains (4--10) $\times 10^4\ M_\odot$ of dust (assuming that the dust temperature is 25 K) and up to $6 \times 10^2$ massive (O5V) stars, appear to be aligned in two parallel ridges, while they have been blended in previous studies.
Despite the similarities in sizes and dust masses of these clumps, their line spectra vary drastically from clump to clump although they are separated by only $\sim 10$ pc.
Specifically, one of the clumps, Clump 1, exhibits line confusion-limited spectra with at least 36 emission lines from 19 molecules (including CH$_3$OH, HNCO, H$_2$CO, CH$_3$CCH, H$_2$CS, and H$_3$O$^+$) and a hydrogen recombination line (H26$\alpha$), while much fewer kinds of molecular lines are detected in some other clumps where fragile species, such as complex organic molecules and HNCO, completely disappear from their spectra. 
We demonstrate the existence of hot molecular gas ($T_\mathrm{rot} (\mathrm{SO_2}) = 90 \pm 11$ K) in the former clump, which suggests that the hot and chemically rich environments are localized within a 10-pc scale star-forming clump.
\end{abstract}

\keywords{galaxies: starburst --- galaxies: ISM --- galaxies: star formation --- galaxies: individual (NGC 253) --- submillimeter: galaxies}

\section{Introduction} \label{sec:intro}
The study of molecular gas in starburst galaxies is essential to understand the diversity of star-formation activities because we can learn the physical properties of molecular gas experiencing an enhanced star-formation rate compared to those in the Milky Way. 
Millimeter/submillimeter (mm/submm) spectroscopic diagnostics are particularly important because intense starbursts are often deeply dust-enshrouded, especially in their early phase where limited information is available at other wavelengths.  
Previous mm/submm spectral scans of local starburst galaxies
(e.g., \citealt{Martin+06}; \citealt{Aladro+11}; \citealt{Martin+11}; \citealt{Costagliola+15}) have demonstrated the richness of the molecular emission lines. 
Spatially resolved spectral line surveys are performed toward a limited number of representative galaxies such as NGC 253 \citep{Meier+15}, NGC 1068 (\citealt{Takano+14}; \citealt{Nakajima+15}), NGC 1097 \citep{Martin+15}, IC 342 \citep{MT05}, and Maffei 2 \citep{MT12}.
However, their spatial resolutions ($\sim 30$--100 pc) are still not sufficient to resolve the individual star-forming regions inside the giant molecular clouds (GMCs). 

High spatial resolution spectroscopic observations are essential to disentangle the multiple physical and chemical processes in the molecular clouds; 
a typical time scale for the chemical evolution of the molecular medium is approximately $10^6$ yrs (e.g., \citealt{Lee+96}), 
whereas the sound crossing time scale for a GMC is approximately $10^7$ yrs \citep{Watanabe+16}.
Therefore, if we observe the GMC-scale molecular properties, 
it will provide an integrated view across the molecular clouds, which is expected to contain multiple components in chemically different stages. 
To resolve the chemical diversity of molecular clouds, or namely,
to discern their evolutionary stages of star-formation with a high time resolution,
the spatially-resolved spectroscopy at clump scales (or smaller) is highly necessary
(e.g., \citealt{Suzuki+92}; \citealt{Sakai+10}; \citealt{Sakai+12}).
Its importance is in fact demonstrated by the recent observations of ST11, 
a high-mass young stellar object (YSO) in the Large Magellanic Cloud (LMC), 
yielding the first extragalactic detection of a hot molecular core, 
one of the early phases of the massive star-formation process \citep{Shimonishi+16}.

Here, we present an 8 pc $\times$ 5 pc resolution view of the central $\sim$ 200 pc region of the nearby starburst galaxy NGC 253, 
based on Atacama Large Millimeter/Submillimeter Array (ALMA) Band 7 observations. 
NGC 253 is one of the nearest prototypical starburst galaxies (e.g. \citealt{Rieke+80});
active star-formation in its central region with the star-formation rate of a few $M_\odot$ yr$^{-1}$ (e.g. \citealt{Radovich+01}) is sustained by a large amount of interstellar medium (e.g., \citealt{Sakamoto+11}).

Throughout the paper, the distance to NGC 253 is adopted to be 3.5 Mpc \citep{Rekola+05}, where $1''$ corresponds to 17 pc.

\section{Observations and Analyses} \label{sec:obs}

We obtain the 340--365 GHz ($\lambda \simeq 0.85$ mm) spectra covering a total frequency range of 11 GHz,
compiling the results of two projects in the ALMA Cycle 2 observations, 
2013.1.00735.S (PI: Nakanishi, hereafter Set-I) and 2013.1.00099.S (PI: Mangum, hereafter Set-II).
The field of view of ALMA at these wavelengths is $16''$--$17''$. 
The Set-I observations cover the frequency range of 340.2--343.4 GHz and 352.5--355.7 GHz, 
and the Set-II observes 350.6--352.4 GHz and 362.2--365.2 GHz. 
These observations were conducted using 34--36 12-m antennas with the baseline lengths 
of 13--784 m (Set-I) and 20--615 m (Set-II).
The uncertainty of the flux calibration in ALMA Band 7 observations is $\sim 10$\% according to the ALMA Cycle 2 Proposer's Guide.

Data reduction is conducted with Common Astronomy Software Applications (CASA; \citealt{McMullin+07}), versions 4.2.1 and 4.2.2.
Using its task \verb|clean|, we image the central region of NGC 253 with Briggs weighting (with a \verb|robust| parameter of 0.5).
We set the same \verb|clean| parameters for the analysis of both datasets to create uniform images over all the frequency ranges.
While we use the data of all $uv$ lengths for each dataset, 
the synthesized beam size of the final images is convolved to $0''.45 \times 0''.3$, 
which corresponds to 8 pc $\times$ 5 pc at the distance to NGC 253. 
With the velocity resolution of 5.0 km s$^{-1}$, 
the rms noise levels are 1.0 and 2.4 mJy beam$^{-1}$ for Set-I and Set-II, respectively.

The continuum map is presented in Figure \ref{fig:images} (a), and it is created in the following manner.
Firstly, we create continuum maps by selecting the channels where no or little line emission appears to exist for four frequency ranges (the upper and lower sidebands of Set-I and Set-II observations, respectively).
Then we take an average weighted according to the widths of the included spectral ranges in the four continuum maps for the purpose of improving the signal-to-noise ratio (S/N).
However, there are some difficulties to define the common line-free channels over the image, since there are a lot of molecular lines with slightly different central velocities on the spectra of different positions.
Therefore, the continuum map above is mostly accurate but not perfect.
Because of this shift in velocities among different positions, the continuum map created on the image may contain $\sim 30$\% uncertainty compared to the continuum determined on the spectra, which is explained in Section \ref{subsec:results1}.
The CS(7--6) line integrated intensity map shown in Figure \ref{fig:images} (b) is created by subtracting the continuum map from the original data cubes on the image-plane.
Although the CS(7--6) line does not have strong neighboring lines, which enables us to make its integrated intensity map,
it is difficult for us to create robust maps for other major lines, which have strong adjacent lines around.

\section{Results} \label{sec:results}

\subsection{Resolved image of the central region of NGC 253} \label{subsec:results1}

\begin{figure*}[p]
\figurenum{1}
\begin{center}
\rotatebox{90}{
\begin{minipage}{\textheight}
\plotone{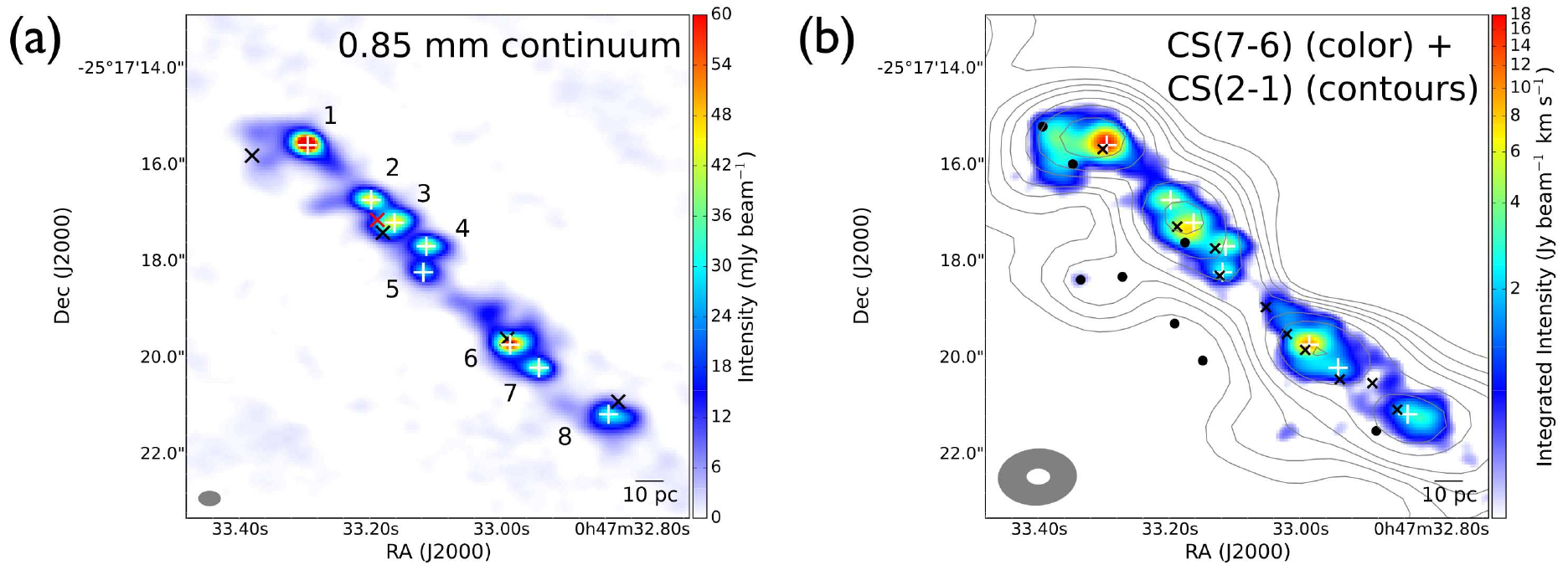}
\centering
\caption{
(a) Band 7 ($\lambda \simeq 0.85$ mm) continuum map. 
White crosses indicate the location of the eight peaks we identify (Clumps 1--8; so as in (b)). 
Red and black cross-marks indicate the positions of the TH02 \citep{TH85} and four out of the five clumps identified by \cite{Sakamoto+11}, respectively. 
The synthesized beam size of this work ($0''.45 \times 0''.3$) is shown at the bottom left corner.
(b) CS(7--6) integrated intensity map, overlaid with the contour of CS(2--1) integrated intensity map obtained by \citet{Meier+15} in ALMA Cycle 0.
The contour levels are 0.5, 1, 1.5, 2, 3, 4, 6, 8, and 10 Jy beam$^{-1}$ km s$^{-1}$.
The synthesized beam sizes of the CS(7--6) ($0''.45 \times 0''.3$; this work) and CS(2--1) ($1''.6 \times 1''.1$; \citealt{Meier+15}) maps are shown at the bottom left corner.
Black symbols indicate the peaks of the centimeter-wave ($\lambda = 1.3$--6 cm) emission sources identified by \cite{UA97};
cross-marks and circles indicate the sources reported as HII region-like and supernova remnant-like objects, respectively.
\label{fig:images}}
\end{minipage}}
\end{center}
\end{figure*}

Figure \ref{fig:images} shows the 0.85 mm continuum map and the CS(7--6) line integrated intensity map.
As shown in Figure \ref{fig:images} (a), the center of NGC 253 is resolved into 10-pc scale star-forming clumps for the first time.  
We identified eight continuum peaks, which are named Clumps 1--8. 
All these clumps are approximately 10 pc in size.
\cite{Sakamoto+11} identified five peaks in their 1.3 mm continuum map with 20 pc ($1''.1$) resolution. 
One of the peaks, peak 3, is resolved into four clumps in our continuum map (Clumps 2--5; see Figure \ref{fig:images} (a)), and it turns out that those have totally diverse spectral features (see Section \ref{subsec:results2}).
In Figure \ref{fig:images} (b), several CS(2--1) peaks \citep{Meier+15} are seen.
In contrast, our CS(7--6) map, which achieves more than three times better angular resolution than that of CS(2--1), resolves the peaks into smaller molecular clumps.

\begin{figure*}[p]
\figurenum{2}
\plotone{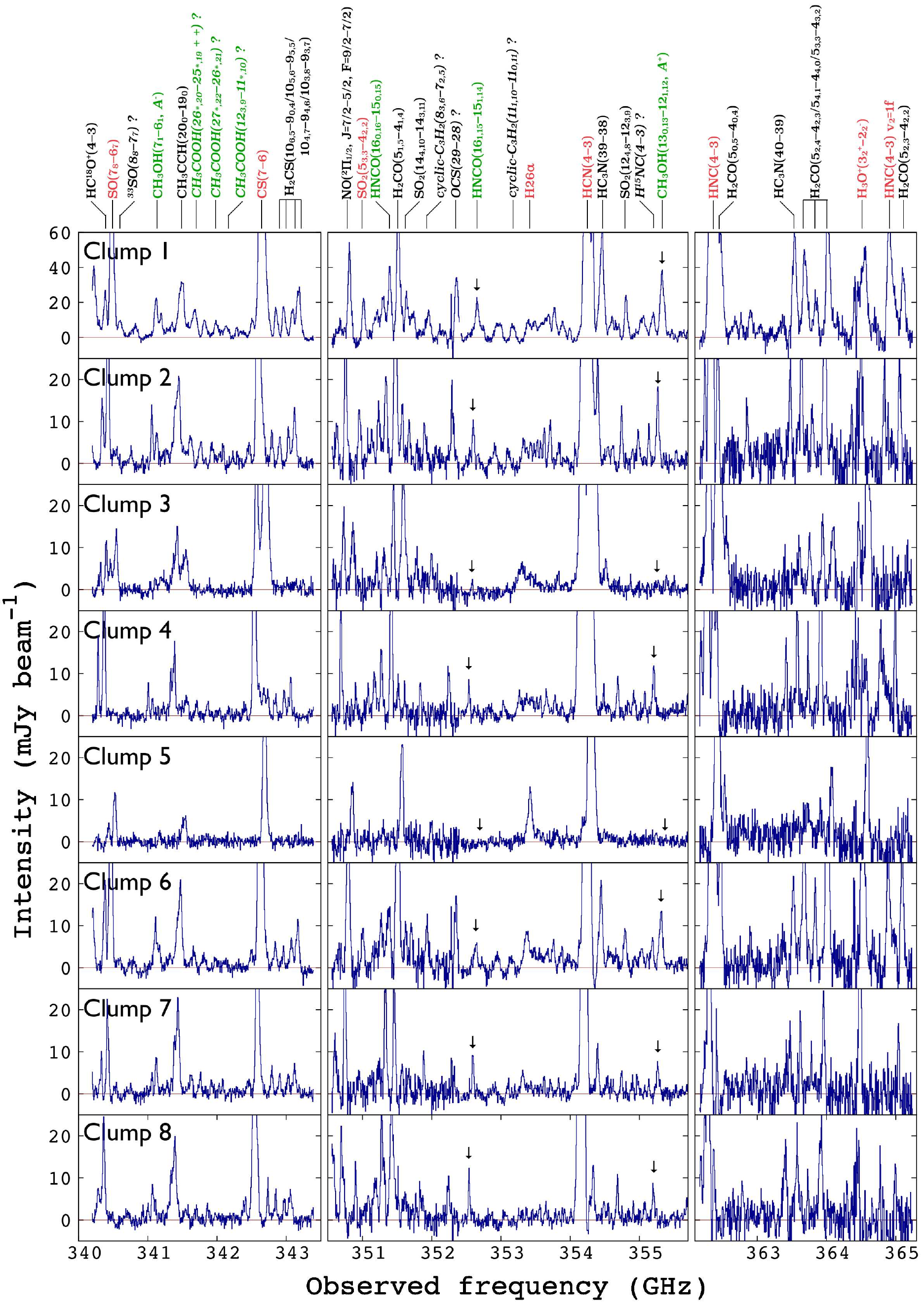}
\caption{Spectra of the eight star-forming clumps we identify. Tentative identifications are labeled in italics. The molecules whose names are written in red and green are ones whose line profiles are shown in Figure \ref{fig:line_profile} and ones known to be easily dissociated by ultraviolet radiation (see Section \ref{subsec:lines}), respectively.
The lines of CH$_3$OH(13$_{0,13}$--12$_{1,12}$, $A^+$) and HNCO(16$_{1,15}$--15$_{1,14}$), which are easily-dissociated molecules and whose spectra are presented in Figure \ref{fig:line_profile}, are indicated by black arrows.
It is probable that NO($^2 \Pi_{1/2}, J=7/2$--$5/2, F=9/2$--$7/2$) line is blended with CH$_3$OH($4_0$--$3_{-1}, E$) line and HC$_3$N(40--39) line with CH$_3$OH($7_2$--$6_1, E$) line.
The rms noise levels are 1.0 mJy beam$^{-1}$ for Set-I (340.2--343.4 GHz and 352.5--355.7 GHz) and 2.4 mJy beam$^{-1}$ for Set-II (350.6--352.4 GHz and 362.2--365.2 GHz), which are too small to be illustrated in the figure above.
\label{fig:spectra_all}}
\end{figure*}

\begin{figure*}[p]
\figurenum{3}
\plotone{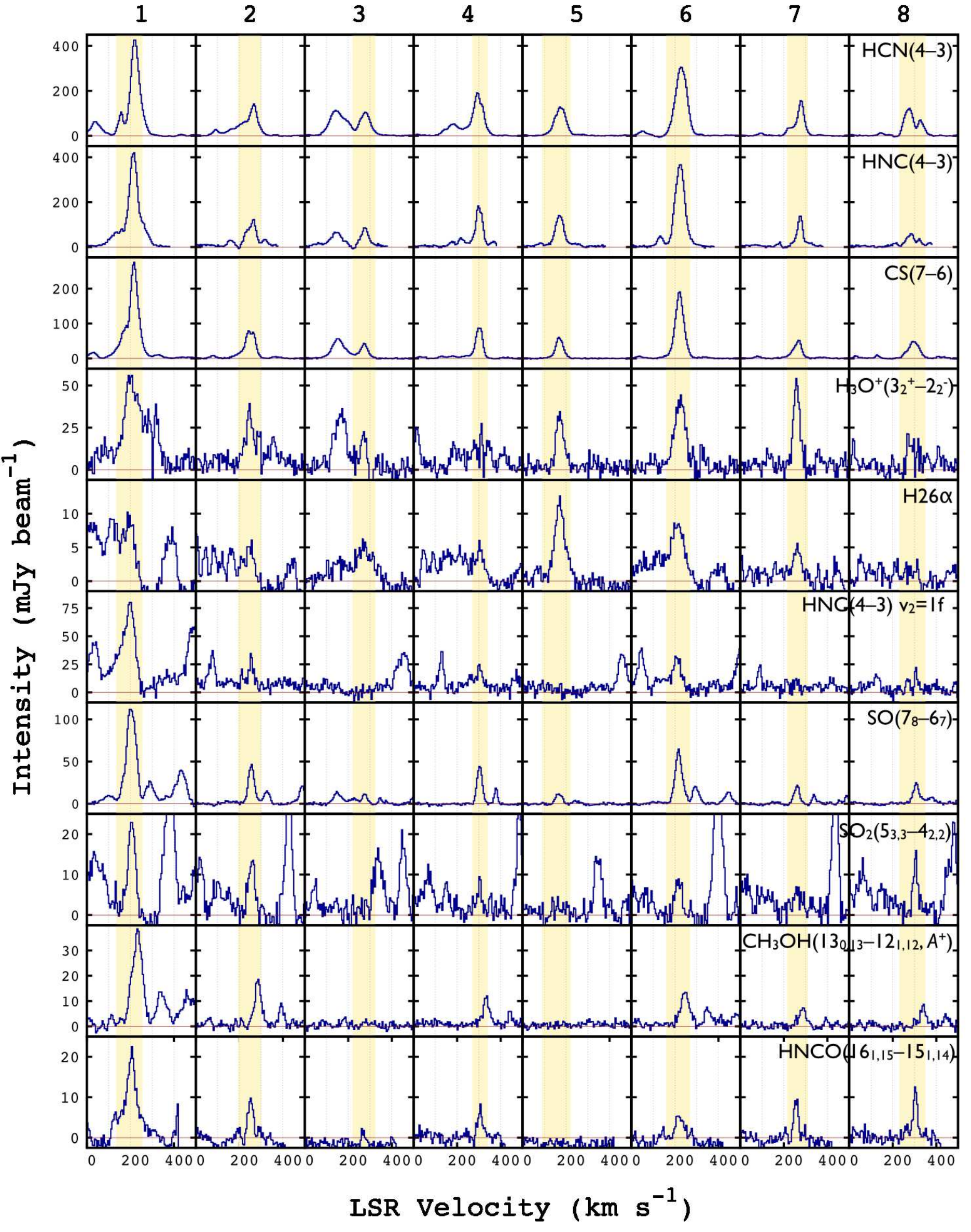}
\caption{Velocity profiles of molecular and recombination lines in Clumps 1--8.
Yellow shaded regions are the velocity ranges we use in deriving the integrated intensity of each clump.
Emission features outside the ranges above are the lines of different species,
except for the low-velocity component in Clump 3. 
\label{fig:line_profile}}
\end{figure*}

We obtain the continuum-subtracted spectra for Clumps 1--8, as shown in Figure \ref{fig:spectra_all}.
The continuum subtraction from the originally-obtained spectra is conducted as follows.
For the four frequency ranges, we measure the continuum intensity of each clump by taking the mean of the values at line-free channels. 
The number of the channels we use to define the continua ranges from 81 out of 560 (the lower sideband of Set-I at Clump 1) to 302 out of 560 (the upper sideband of Set-I at Clump 5).
These channels are located over the whole frequency range we analyze; some are taken from just several channels at the narrow valleys between emission lines, which are placed at intervals of typically $\sim 0.2$ GHz. 
Differing the channels for respective clumps enables us to precisely determine the continuum level for each clump.
Then we subtract the continua from the spectra of Clumps 1--8 for the four spectral ranges, which results in the continuum-subtracted spectra of eight clumps shown in Figure \ref{fig:spectra_all}.
Because the entire profiles of the high intensity lines cannot fit in Figure \ref{fig:spectra_all}, they are shown in Figure \ref{fig:line_profile}.

In terms of the spatial distributions of the clumps,
Clumps 1--4 and 5--8 are aligned in two parallel ridges (hereafter Ridge-N and Ridge-S), respectively.
These ridges are approximately 5 pc apart in the projected distance.
As shown in Figure \ref{fig:line_profile}, the line peak velocity increases on each ridge, from the northeast to the southwest,
which is likely to correspond to the orbital motion around the galactic center.
We consider that the galactic center is located at TH02 (R.A.(J2000) $= 0^\mathrm{h} 47^\mathrm{m} 33^\mathrm{s}.2$, Dec.(J2000) $= -25^\circ 17'16''.9$; \citealt{TH85}), the brightest centimeter-wave continuum source of thermal emission.
The spectra in Clump 3 exhibit a double velocity component;
it is possible that the high and low-velocity components are due to the contributions from Ridge-N and Ridge-S, respectively.
There appear to be double-peak features of self-absorption in the profiles of presumably optically thick HCN and HNC lines at Clump 8, while such features are absent in those of optically thin lines such as SO and HNCO.

\begin{table*}[p]
\begin{center}
\rotatebox{90}{
\begin{minipage}{\textheight}
\centering
  \caption{Emission parameters of the eight star-forming clumps.}
  \begin{tabular}{llrrrrrrrr}
\hline
\multicolumn{2}{l}{Clump}	&	1	&	2	&	3	&	4	&	5	&	6	&	7	&	8	  \\  \hline
Peak position & R.A.($0^\mathrm{h} 47^\mathrm{m}$) &$	33	^\mathrm{s}.	30	$&$	33	^\mathrm{s}.	16	$&$	33	^\mathrm{s}.	20	$&$	33	^\mathrm{s}.	11	$&$	33	^\mathrm{s}.	12	$&$	32	^\mathrm{s}.	99	$&$	32	^\mathrm{s}.	94	$&$	32	^\mathrm{s}.	84	$\\ 
(J2000) & Dec.($-25^\circ 17'$) &$	15	''.	6	$&$	17	''.	2	$&$	16	''.	7	$&$	17	''.	7	$&$	18	''.	2	$&$	19	''.	7	$&$	20	''.	2	$&$	21	''.	2	$ \\ \hline
\multicolumn{2}{l}{Size\tablenotemark{a} (pc)}	&	$8.8 \times 6.0 $	&	$10.5 \times 6.0$		&	$11.2 \times 8.0$		&	$8.3 \times 5.3$	&	$9.9 \times 7.6 $		&	$9.2  \times 5.8$		&	$9.3 \times 5.7 $		&	$15.0 \times 8.6$		  \\  \hline
\multicolumn{2}{l}{Velocity range\tablenotemark{b} (km s$^{-1}$)}	&	135--255	&	195--300	&	220--325	&	270--340	&	90--220	&	160--270	&	215--310	&	230--350	\\ \hline
&HCN(4--3)	&$	23.40	\pm	0.06	$&$	8.34	\pm	0.07	$&$	6.38	\pm	0.03	$&$	9.03	\pm	0.03	$&$	7.94	\pm	0.05	$&$	19.32	\pm	0.06	$&$	6.83	\pm	0.04	$&$	8.17	\pm	0.05	$ \\	
&HNC(4--3)	&$	23.73	\pm	0.08	$&$	6.09	\pm	0.20	$&$	3.83	\pm	0.13	$&$	6.95	\pm	0.13	$&$	7.37	\pm	0.12	$&$	20.01	\pm	0.08	$&$	4.85	\pm	0.11	$&$	3.53	\pm	0.14	$ \\	
&CS(7--6)	&$	15.20	\pm	0.09	$&$	4.55	\pm	0.04	$&$	2.01	\pm	0.03	$&$	3.40	\pm	0.04	$&$	2.65	\pm	0.02	$&$	9.25	\pm	0.03	$&$	2.33	\pm	0.03	$&$	2.99	\pm	0.06	$ \\	
&H$_3$O$^+$($3_2^+$--$2_2^-$)	&$	4.50	\pm	0.31	$&$	2.01	\pm	0.13	$&$	0.64	\pm	0.08	$&$	0.93	\pm	0.13	$&$	1.51	\pm	0.11	$&$	2.57	\pm	0.21	$&$	1.89	\pm	0.10	$&$	0.90	\pm	0.11	$ \\	
Integrated intensity\tablenotemark{c}&H26$\alpha$ 	&$	0.66	\pm	0.08	$&$	0.29	\pm	0.05	$&$	0.40	\pm	0.02	$&$	0.24	\pm	0.04	$&$	0.68	\pm	0.03	$&$	0.62	\pm	0.03	$&$	0.25	\pm	0.03	$&$<	0.09	$  \\			
(Jy beam$^{-1}$ km s$^{-1}$)&HNC(4--3) $v_2=1f$	&$	5.49	\pm	0.14	$&$	1.46	\pm	0.13	$&$<	0.19	$ &$			0.97	\pm	0.07	$&$<	0.27	$ &$			1.71	\pm	0.14	$&$<	0.22	$ &$<			0.33	$  \\			
&SO($7_8$--$6_7$)	&$	6.35	\pm	0.08	$&$	1.64	\pm	0.12	$&$	0.47	\pm	0.05	$&$	1.31	\pm	0.11	$&$	0.41	\pm	0.03	$&$	2.62	\pm	0.12	$&$	0.62	\pm	0.05	$&$	0.94	\pm	0.07	$ \\	
&SO$_2$($5_{3,3}$--$4_{2,2}$)	&$	1.03	\pm	0.07	$&$	0.52	\pm	0.05	$&$<	0.14	$ &$			0.26	\pm	0.04	$&$<	0.15	$ &$			0.30	\pm	0.05	$&$	0.33	\pm	0.04	$&$	0.47	\pm	0.06	$ \\ 
&CH$_3$OH(13$_{0,13}$--12$_{1,12}$, $A^+$)	&$	2.01	\pm	0.08	$&$	0.66	\pm	0.04	$&$<	0.07	$&$	0.37	\pm	0.02	$&$<	0.08	$&$	0.65	\pm	0.05	$&$	0.28	\pm	0.02	$&$	0.25	\pm	0.03	$ \\
&HNCO(16$_{1,15}$--15$_{1,14}$)	&$	1.26	\pm	0.04	$&$	0.31	\pm	0.02	$&$<	0.06	$&$	0.19	\pm	0.02	$&$<	0.09	$&$	0.25	\pm	0.03	$&$	0.24	\pm	0.02	$&$	0.37	\pm	0.03	$ \\ \hline
\multicolumn{2}{l}{FWHM of CS(7--6) line (km s$^{-1}$)}	&$46\pm1$&$54\pm2$&$43\pm1$&$36\pm1$&$41\pm1$&$47\pm1$&$43\pm2$&$56\pm1$ \\
\hline 
\multicolumn{2}{l}{Continuum intensity at the peak\tablenotemark{d} (mJy beam$^{-1}$)}	&$	81 \pm 2	$&$47	\pm	2	$&$	46	\pm	2	$&$	40	\pm	2	$&$	26	\pm	2	$&$	61	\pm	2	$&$	38	\pm	2	$&$	30\pm	2	$ \\ \hline 
\multicolumn{2}{l}{Continuum flux density over the clump size\tablenotemark{d}  (mJy)}	&$	106 \pm 3	$&$71	\pm	3	$&$	92	\pm	2	$&$	37	\pm	2	$&$	49	\pm	2	$&$	70	\pm	2	$&$	42	\pm	2	$&$	102\pm	4	$ \\ \hline 
\multicolumn{2}{l}{$M_{\mathrm{dust}}$\tablenotemark{e} ($M_\odot$)} & $1 \times 10^5$ & $7 \times 10^4$ & $9 \times 10^4$ & $4 \times 10^4$ & $5 \times 10^4$ & $7 \times 10^4$ & $4 \times 10^4$ & $1 \times 10^5$ \\ \hline
\multicolumn{2}{l}{Number of O5V stars\tablenotemark{f}} & $6 \times 10^2$ & $3 \times 10^2$ & $4 \times 10^2$ & $2 \times 10^2$ & $6 \times 10^2$ & $6 \times 10^2$ & $2 \times 10^2$ & $<8 \times 10^1$ \\ \hline
  \end{tabular}
  \tablenotetext{a}{\raggedright Clump sizes (FWHM) are determined by the 2-dimensional Gaussian fitting of the continuum map. Although the uncertainties of the fitting are about 1\% of the sizes, the sizes above are just rough estimates, since it is in principle difficult to clearly define the emitting region of each clump.}
  \tablenotetext{b}{\raggedright Line LSR velocity ranges we use in integrating the continuum-subtracted spectra shown in Figure \ref{fig:line_profile}.
  Note that we integrate only the high-velocity component (originated from Ridge-N) in Clump 3, considering the line profile of H26$\alpha$, while the integration is conducted over the line profiles in the other clumps, including Clumps 1 and 8, where the features of self-absorption appear to exist.}
  \tablenotetext{c}{\raggedright 
  We integrate the lines spatially over the beam at the continuum peak positions, and spectrally over the velocity ranges above.
  Errors are derived only from the rms noise levels of continuum channels, which are located on one or both sides of each line profile on the spectra.
  Note that the integrated intensities of H26$\alpha$ lines have potentially large uncertainties, since the H26$\alpha$ lines possibly have non-identical profiles to other molecular lines to some extent and we cannot exclude the possibility that they are blended with minor neighboring molecular lines.
  For non-detected lines, the $3\sigma$ upper limits of the integrated intensities are shown.}
  \tablenotetext{d}{\raggedright 
  We derive both the continuum intensities over the beam at the peak positions and the continuum flux density over the clump sizes, each of which is averaged over our 11 GHz band.
  }
  \tablenotetext{e}{\raggedright 
  As dust masses are calculated from the continuum flux densities over the clump sizes, their errors due to the S/N are about several percents and therefore negligible when the masses are rounded to one significant number.
  Note that the values are rough estimate, since the dust continuum fluxes might vary by at most $\sim 30$\% over the frequency range we analyze, and the continuum emission is assumed to be purely composed of the dust thermal emission.
  }
  \tablenotetext{f}{\raggedright 
  The numbers of O5V stars are just rough estimates to compare each clump to other, since they are derived from the H26$\alpha$ integrated intensities, which contain potentially large uncertainties (see Note {\it c}). 
  The numbers of O5V stars decrease to one-third of the values shown above, if we adopt the value of $N_L$ given by \cite{Panagia73} instead of that given by \cite{Martins+05}.
  }
  \label{tab1}
  \end{minipage}}
\end{center}
\end{table*}

\subsection{Drastic difference of spectral features among 10-pc scale star-forming clumps}
\label{subsec:results2}
In the spectra of the continuum-selected Clumps 1--8, we identify 37 lines originating from 19 molecular and one atomic species, including several tentative identifications.
The line identification is conducted by comparing the spectra we obtained with the molecular line database Splatalogue\footnote{http://www.splatalogue.net/}, which mainly contains data from the Cologne Database for Molecular Spectroscopy \citep{Muller+05} and the Jet Propulsion Laboratory catalogues \citep{Pickett+98}.
We also refer to the results of the line surveys toward Galactic star-forming regions with similar frequency ranges (\citealt{Sutton+91}; \citealt{HvD97}; \citealt{Schilke+97}; \citealt{Nagy+15}).
We identify the lines by preferentially choosing the molecules and their transitions that are previously detected in local star-forming regions.
How well the lines are separated in each clump is determined mainly by its line widths, not by the instrumental resolution.
Although further observations in different frequency ranges and the comparison with the other transitions of the detected lines are needed for the definitive line identification, we conduct the best identification by eye that could be done with the current data.

Surprisingly, the molecular line appearances and intensities drastically differ from clump to clump,
even though they are each separated by approximately 10 pc in projection (Figure \ref{fig:spectra_all}).
Although the continuum intensities vary by at most $\sim 3$ times among the clumps, the variety of the apparent patterns of their spectra is more prominent than the difference of the S/N.
Clump 1 exhibits line confusion-limited spectra with many lines of various molecular species, 
while a much smaller number of molecular lines are detected in Clumps 3 and 5.
Clumps 2, 4, 6, 7, and 8 appear to be classified as an intermediate-type; 
their spectra do not reach the line confusion-limit, while a larger number of molecules are detected therein than in Clumps 3 and 5.
Now we resolve and distinguish the individual features of the 10-pc scale star-forming clumps in an extragalactic source,
which have been blended and averaged over several tens-of-parsecs in scale in previous studies (e.g., \citealt{Sakamoto+11}; \citealt{Meier+15}).
Although the chemical diversities over parsec scales have been reported in Galactic sources such as Sgr B2 (e.g., \citealt{Sutton+91}) and W3 (e.g., \citealt{HvD97}),
our work resolves the diverse nature outside the Local Group for the first time.
 detailed discussions about the chemical variety of the clumps based on the their line ratios are given in Section \ref{subsec:lines}.

\subsection{Line profiles and emission parameters}
\label{subsec:results3}
Line profiles of nine major molecular lines and hydrogen recombination line H26$\alpha$ are shown in Figure \ref{fig:line_profile}.
Table \ref{tab1} summarizes the integrated intensities of these ten lines and the continuum intensities of Clumps 1--8.
We integrate the line emission spatially over the beam at the continuum peak positions, and spectrally over the velocity ranges we set for respective clumps, because of the following reason.
As the spectra are close to be line confusion-limited, it is virtually impossible to set the common velocity ranges to integrate the molecular lines concerned without blending the neighboring ones.
Therefore, we set the different frequency ranges for each clump, over which we spectrally integrate the lines, instead of creating the integrated intensity maps for every molecular line with the same velocity ranges.
If an integrated intensity map of a certain line is created with the particular frequency range that covers the whole line profile at all the clumps, for example, 90--350 km s$^{-1}$, it is inevitable that different lines are blended at some of the clumps and the integrated intensity of the line concerned are overestimated.
Since we cannot determine the line emitting regions robustly without integrated intensity maps, we spatially integrate the lines over the beam as point sources at the continuum peak positions, so that we can directly compare the integrated intensity of a clump with those of other ones.

We do not fit the lines with Gaussian profiles, since their unsymmetrical profiles and the existence of a lot of their neighboring lines prevent the precise Gaussian fitting.
Since the rms errors of the integrated intensities shown in Table \ref{tab1} arise only from S/N of the lines, there may be larger uncertainties especially for the lines such as H26$\alpha$ with profiles far apart from Gaussian.
These values should be directly compared, since we adjust the way integrate the lines for all the species and clumps, despite the possible uncertainties of the integrated intensities.

The dust masses ($M_{\mathrm{dust}}$) of the clumps are calculated from the continuum flux density with the following standard equation: 
\begin{equation}
M_{\mathrm{dust}} = \frac{S_\nu D^2}{\kappa_\nu B_\nu(T_{\mathrm{dust}})}
\end{equation}
(e.g., \citealt{Klaas+01}; \citealt{Krips+16}), where $S_\nu$ is the flux observed at frequency $\nu$,
$D$ is the distance to the source, $\kappa_\nu$ is the mass absorption coefficient,
and $B_\nu(T_{\mathrm{dust}})$ is the Planck function 
at frequency $\nu$ and dust temperature $T_{\mathrm{dust}}$.
Here, we adopt the averaged continuum flux density over the clump size as $S_\nu$,
$\nu = 353$ GHz (weighted averaged value of the whole frequency range analyzed),
$D = 3.5$ Mpc \citep{Rekola+05},
and $\kappa_\nu = 0.0865$ m$^2$ kg$^{-1}$ at $\nu = 353$ GHz \citep{Klaas+01}.
\cite{Krips+16} derived the dust temperature for the NGC 253 disk: $T_{\mathrm{dust}} = 25$ K.
We adopt this value, although $M_{\mathrm{dust}}$ decreases by a factor of 20 when $T_{\mathrm{dust}}$ varies from 10 K to 100 K. 
Note that the derived values of dust masses are rough estimates, because the dust continuum fluxes might vary by $\sim 30$\% when the value at the lower edge of the frequency range is compared to that at the upper edge. 
Nevertheless, the relative comparison of the masses among the clumps is more reliable, since the procedure to derive the masses is common among the clumps. 

Although here we assume that the dust thermal emission comprises all of the continuum emission, we actually find minor contribution from free-free emission.
The contribution could account for up to 20\% of 353 GHz continuum flux density, which is estimated by extrapolating the 99 GHz free-free flux density given by \cite{Bendo+15}.
Note that, however, this estimate is just an upper-limit, since the observations conducted by \cite{Bendo+15} covered the emission from larger spatial scale than we do, as the minimum baseline length of their observations (7.0k$\lambda$; \citealt{Bendo+15}) is half as long as that of ours (15.3k$\lambda$).
Although the estimate of dust masses of the clumps presented in Table \ref{tab1} needs to be slightly reduced, the correction up to 20\% does not affect the comparison of the scales of the clumps.

The number of O5V stars of each clump is derived from its integrated intensity of the H26$\alpha$ line.
We assume all the ionizing photons arise from a single stellar population (O5V) and they are perfectly absorbed by ambient neutral hydrogen.
H26$\alpha$ integrated intensities are converted into Lyman continuum
photon fluxes (s$^{-1}$) in the following manner. We estimate the equivalent H$\beta$ flux from the observed H26$\alpha$ flux by using the line ratio presented in \cite{HS87}. Then, Lyman continuum photon number is calculated by using the H$\beta$ flux the recombination coefficients presented by \cite{Osterbrock89}.
We assume that the electron density and temperature of ionized gas are $10^3 \ {\rm cm}^{-3}$ and $10^4 \ {\rm K}$, respectively (Nakanishi et al. in preparation), and Case B condition \citep{Osterbrock89}.
The numbers of O5V stars in individual clumps are obtained by dividing the Lyman continuum photon fluxes by the Lyman continuum photon flux per single O5V star ($N_L$).
We adopt $\log{N_L} = 49.26$ given by \cite{Martins+05}.
The number of O5V stars decrease to one-third of the values shown in Table \ref{tab1} if we adopt the theoretical value $\log{N_L} = 49.71$ given by \cite{Panagia73}, which has little influence upon the following discussions.

\begin{figure}[t!]
\figurenum{4}
\plotone{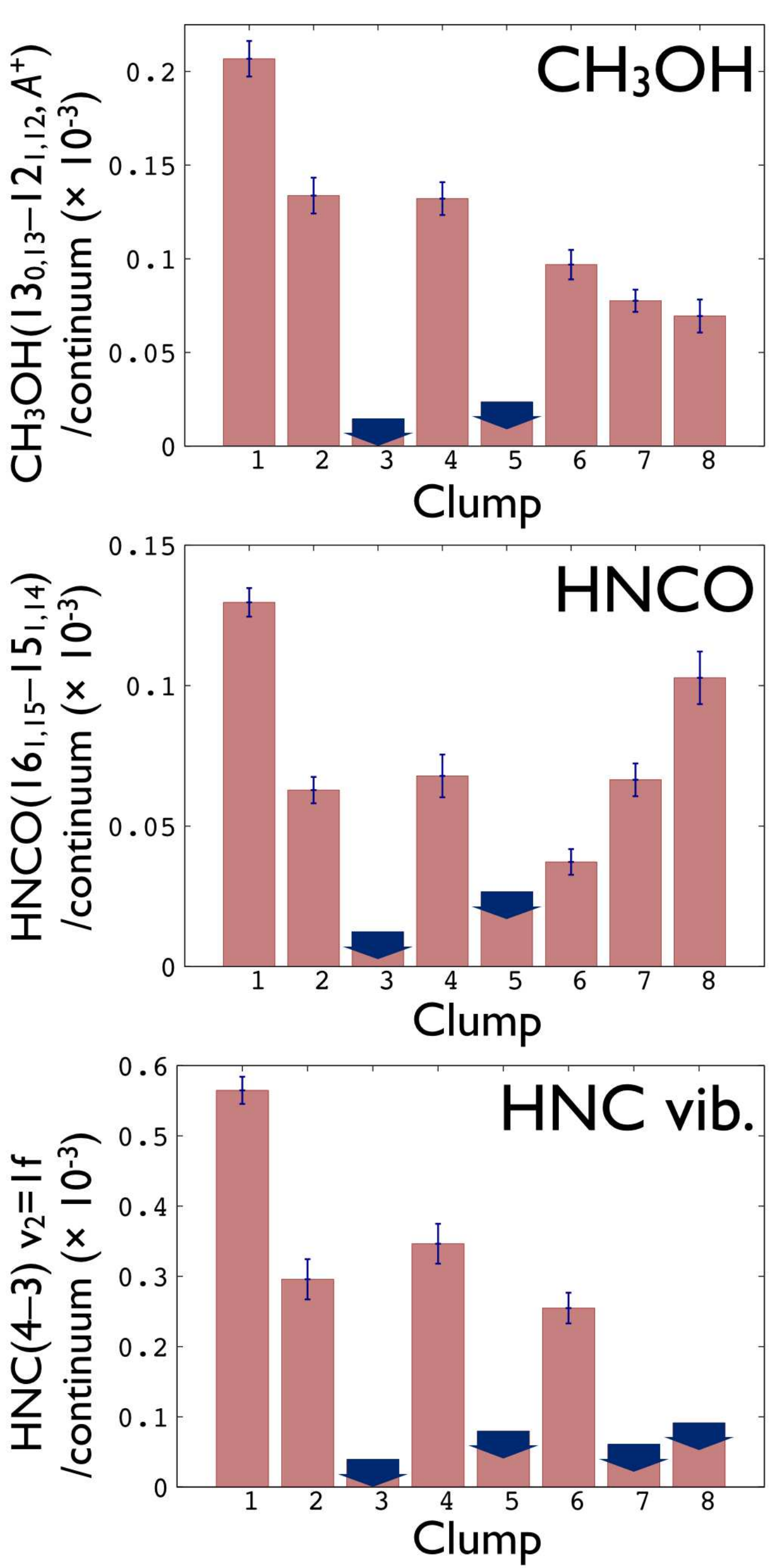}
\caption{Integrated intensity ratios of CH$_3$OH(13$_{0,13}$--12$_{1,12}$, $A^+$), HNCO(16$_{1,15}$--15$_{1,14}$), and the HNC vibrationally excited line ($J = 4$--3, $v_2 = 1f$) to continuum for Clumps 1--8. For non-detected lines, the $3\sigma$ upper limits of the ratios are shown with blue arrows.
\label{fig:line_ratios}}
\end{figure}

The integrated intensity ratios of the ten representative lines to continuum are summarized in Table \ref{tab2}.
We derive the ratios by dividing the the integrated intensities of the lines shown in Table \ref{tab1} by the integrated continuum intensities, which are continuum intensities at the eight peak positions (see Table \ref{tab1}) spectrally integrated over the same velocity ranges as those used when integrating the lines for each clump.
The integrated intensity ratios of the three lines (CH$_3$OH(13$_{0,13}$--12$_{1,12}$, $A^+$), HNCO(16$_{1,15}$--15$_{1,14}$), and HNC(4--3) $v_2=1f$) to continuum are illustrated in Figure \ref{fig:line_ratios}.

\begin{table*}[p]
\begin{center}
\rotatebox{90}{
\begin{minipage}{\textheight}
\centering
  \caption{Integrated intensity ratios of ten representative lines to continuum for the eight star-forming clumps.}
  \begin{tabular}{llrrrrrrrr}
\hline
\multicolumn{2}{l}{Clump}	&	1	&	2	&	3	&	4	&	5	&	6	&	7	&	8	  \\  \hline
\multicolumn{2}{l}{Velocity range (km s$^{-1}$)}	&	135--255	&	195--300	&	220--325	&	270--340	&	90--220	&	160--270	&	215--310	&	230--350	\\ \hline
&HCN(4--3)	&$	2.41	\pm	0.06	$&$	1.69	\pm	0.07	$&$	1.32	\pm	0.03	$&$	3.23	\pm	0.12	$&$	2.35	\pm	0.09	$&$	2.88	\pm	0.07	$&$	1.89	\pm	0.05	$&$	2.27	\pm	0.10	$ \\
&HNC(4--3)	&$	2.44	\pm	0.06	$&$	1.23	\pm	0.06	$&$	0.79	\pm	0.03	$&$	2.48	\pm	0.11	$&$	2.18	\pm	0.09	$&$	2.98	\pm	0.08	$&$	1.34	\pm	0.05	$&$	0.98	\pm	0.06	$ \\
&CS(7--6)	&$	1.56	\pm	0.04	$&$	0.92	\pm	0.04	$&$	0.42	\pm	0.01	$&$	1.21	\pm	0.05	$&$	0.78	\pm	0.03	$&$	1.38	\pm	0.04	$&$	0.65	\pm	0.02	$&$	0.83	\pm	0.04	$ \\
&H$_3$O$^+$($3_2^+$--$2_2^-$)	&$	0.46	\pm	0.03	$&$	0.41	\pm	0.03	$&$	0.13	\pm	0.02	$&$	0.33	\pm	0.05	$&$	0.45	\pm	0.04	$&$	0.38	\pm	0.03	$&$	0.52	\pm	0.03	$&$	0.25	\pm	0.03	$ \\
Integrated intensity ratio&H26$\alpha$	&$	0.07	\pm	0.01	$&$	0.06	\pm	0.01	$&$	0.08	\pm	0.01	$&$	0.09	\pm	0.01	$&$	0.20	\pm	0.01	$&$	0.09	\pm	0.01	$&$	0.07	\pm	0.01	$&$<	0.03	$\tablenotemark{b} \\		
to continuum\tablenotemark{a} ($\times 10^{-3}$)&HNC(4--3) $v_2=1f$	&$	0.56	\pm	0.02	$&$	0.30	\pm	0.03	$&$<	0.04	$\tablenotemark{b}&$			0.35	\pm	0.03	$&$<	0.08	$\tablenotemark{b}&$			0.25	\pm	0.02	$&$<	0.06	$\tablenotemark{b}&$<			0.09	$\tablenotemark{b} \\	
&SO($7_8$--$6_7$)	&$	0.65	\pm	0.02	$&$	0.33	\pm	0.03	$&$	0.10	\pm	0.01	$&$	0.47	\pm	0.04	$&$	0.12	\pm	0.01	$&$	0.39	\pm	0.02	$&$	0.17	\pm	0.01	$&$	0.26	\pm	0.02	$ \\
&SO$_2$($5_{3,3}$--$4_{2,2}$)	&$	0.11	\pm	0.01	$&$	0.11	\pm	0.01	$&$<	0.03	$\tablenotemark{b}&$			0.09	\pm	0.01	$&$<	0.04	$\tablenotemark{b}&$			0.04	\pm	0.01	$&$	0.09	\pm	0.01	$&$	0.13	\pm	0.02	$ \\
&CH$_3$OH(13$_{0,13}$--12$_{1,12}$, $A^+$)	&$	0.21	\pm	0.01	$&$	0.13	\pm	0.01	$&$<	0.01	$\tablenotemark{b}&$			0.13	\pm	0.01	$&$<	0.02	$\tablenotemark{b}&$			0.10	\pm	0.01	$&$	0.08	\pm	0.01	$&$	0.07	\pm	0.01	$ \\
&HNCO(16$_{1,15}$--15$_{1,14}$)	&$	0.13	\pm	0.01	$&$	0.06	\pm	0.01	$&$<	0.01	$\tablenotemark{b}&$			0.07	\pm	0.01	$&$<	0.03	$\tablenotemark{b}&$			0.04	\pm	0.01	$&$	0.07	\pm	0.01	$&$	0.10	\pm	0.01	$ \\ \hline
  \end{tabular}
  \tablenotetext{a}{\raggedright The integrated intensity ratios are derived as follows; the integrated intensities of the lines shown in Table \ref{tab1} are divided by the continuum intensities at the eight peak positions integrated over the same velocity ranges as those used in the integration of the lines for each clump. The errors are propagated from the rms errors of the integrated intensities of the lines and the continuum intensities. For the simplicity, the ratios shown are multiplied by $10^3$.}
  \tablenotetext{b}{\raggedright For non-detected lines, the $3\sigma$ upper limits of the integrated intensity ratios are shown.}
  \label{tab2}
  \end{minipage}}
\end{center}
\end{table*}

\section{Discussion} \label{sec:discussion}

\subsection{Physical similarity and chemical diversity of star-forming clumps} \label{subsec:lines}
The features of the molecular line spectra are notably different from clump to clump, despite their physical similarity; the sizes and line width of the clumps vary only by twice.
Although the variations of their dust masses and the number of O5V stars are relatively large (five times or less), they are surpassed by their distinguishing chemical diversity.

The most striking chemical difference between clumps is the number of detected molecules.
As shown in Figure \ref{fig:spectra_all},
Clump 1 exhibits line confusion-limited spectra where their continua are not easily identified, 
while much fewer kinds of molecular lines are detected in Clumps 3 and 5.
In the latter clumps, 
molecular lines as a whole seem to be suppressed, which is obvious even when compared to the clumps which have similar continuum intensities, such as Clump 8.
Especially, HNCO lines (HNCO($16_{0,16}$--$15_{0,15}$) and HNCO($16_{1,15}$--$15_{1,14}$)) and CH$_3$OH lines (CH$_3$OH($7_1$--$6_1$, $A^-$) and CH$_3$OH(13$_{0,13}$--12$_{1,12}$, $A^+$)) completely disappear from the spectra of Clump 5, while they are clearly detected in Clump 8.
The integrated intensity ratio of HNCO($16_{1,15}$--$15_{1,14}$) to continuum, presented in Table \ref{tab2} and Figure \ref{fig:line_ratios}, is $(0.10 \pm 0.01) \times 10^{-3}$ in Clump 8, while it falls to below $0.03 \times 10^{-3}$ in Clump 5.
As HNCO tends to be dissociated by ultraviolet (UV) photons 
(e.g., \citealt{Martin+08}; \citealt{Martin+09b}), 
it is probable that HNCO is destroyed by strong UV radiation from massive stars, particularly in Clump 5, which is the one with the highest number of O5V stars.
Other clumps than Clumps 3 and 5, however, can harbor shielded regions where HNCO can survive in comparably intense UV radiation.

CH$_3$OH also appears to be suppressed in Clumps 3 and 5; 
for example, the integrated intensity ratio of CH$_3$OH(13$_{0,13}$--12$_{1,12}$, $A^+$) to continuum in Clump 5 is less than $0.02 \times 10^{-3}$, while it takes $(0.07 \pm 0.01) \times 10^{-3}$ even in comparably-bright Clump 8, and it takes as high as $(0.21 \pm 0.01) \times 10^{-3}$ in Clump 1.
CH$_3$OH, one of the complex organic molecules (COMs; \citealt{HvD09}), is also easily dissociated by cosmic rays and UV photons (\citealt{Martin+06}; \citealt{Martin+09b}; \citealt{Aladro+13}).
This additionally supports the fact that some shielding mechanisms work in Clump 1 and others except Clumps 3 and 5.
Furthermore, CH$_3$COOH, an even larger COM tentatively identified in Clump 1, also disappears in Clumps 3 and 5.
The suppression of molecular lines, especially those of ones easily dissociated by UV photons, suggests that Clumps 3 and 5 are filled with HII regions with hundreds of O-type stars.

Note that, however, there are some other scenarios accounting for the suppression of HNCO and COMs.
One possibility is that such molecules have not produced on dust grains yet, or at least have not sublimed into gas-phase.
This scenario is consistent with that sulfur-bearing species such as SO and SO$_2$ are also suppressed in Clumps 3 and 5, since the formation of such sulfur-bearing species is largely determined during the phase when atomic sulfur freezes out on dust grains \citep{Wakelam+04}.
Another scenario that can explain the molecular gas suppression is that the fraction of line-emitting cores are relatively low in Clumps 3 and 5.
We cannot resolve a single star-forming core with the current observations, and it is possible that the lines from the molecules prone to concentrate around the core instead of existing over ambient diffuse gases are diluted over a 10-pc scale beam.
HNCO, SO, and SO$_2$ seem to tend to distribute compactly around a star-forming core (e.g. \citealt{Nagy+15}).

Previous studies have found the specific spatial distributions of HNCO and/or CH$_3$OH gases inside nearby galaxies (e.g. \citealt{MT05}; \citealt{MT12}; \citealt{Martin+15}; \citealt{Saito+17}; \citealt{Ueda+17}; \citealt{Tosaki+17}), and the distribution of warm molecular gases in NGC 253 is also investigated (e.g. \citealt{Ott+05}; \citealt{Krips+16}; \citealt{Gorski+17}).
Nevertheless, the spatial resolution of these observations are limited to $\sim$ 30--100 pc, and the drastic chemical diversity among even smaller 10-pc scale clumps has not been unveiled until this work.

The distribution of the HNC vibrationally excited line ($J = 4$--3, $v_2 = 1f$) emission is also noticeable.
We detect the line in Clumps 1, 2, 4, and 6, which is remarkable in that it is for the first time that the vibrationally excited line of HNC is detected in NGC 253.
This is the third detection in external galaxies after the ones in the AGN-hosting luminous infrared galaxies NGC 4418 (\citealt{Costagliola+13}; \citealt{Costagliola+15}) and IRAS 20551--4250 \citep{INI16}.
While the vibrationally excited line is absent in Clumps 3, 5, 7, and 8,
its emission is outstandingly enhanced in Clump 1;
the intensity ratio of the vibrationally excited line to the continuum is $(0.56 \pm 0.02) \times 10^{-3}$ in Clump 1,
which is much higher than those in the other clumps ($< 0.35 \times 10^{-3}$).
The HNC molecule, which absorbs mid-infrared photons with a wavelength of $\lambda = 21.5$ $\mu$m and an energy level of $h\nu/k = 669$ K, decays back to its ground state after being pumped up to its bending vibrational mode (infrared pumping; \citealt{Aalto+07}).
The enhancement of the HNC vibrationally excited line in Clump 1 suggests the existence of mid-infrared radiation sources and that active infrared pumping takes place therein.
However, the HNC vibrationally excited line intensity does not simply correlate with the {\it Q}-band ($\lambda = 18.72$ $\mu$m) continuum intensity presented by \cite{FO+09}; {\it Q}-band emission appears to be brighter around Clump 6 than that around Clump 1, where the HNC vibrationally excited line intensity is largest among the clumps.
We might have to consider not only infrared pumping but also some other mechanisms as the reason for the prominent emission of the vibrationally excited line in Clump 1.

We consider that the internal velocity field does not dominantly affect the apparent spectral features and the line ratios. 
The line widths of the molecular lines have little variance (by 1.6 times at most) among the clumps (see Table \ref{tab1}).
Although there seem to be no prominent difference of spectral features in most of the line profiles presented in Figure \ref{fig:line_profile}, some of the clumps exhibit complex line profiles, which could be attributed to the spatial structures of molecular gases therein.
For example, Clump 1 appears to have wing-like line profile, which possibly imply that Clump 1 is more extended in projection and along the line-of-sight spatially.
Nevertheless, the contribution from the wing-like part to the total line flux is just limited (e.g. 14\% in its CS integrated intensity).
In addition, there seem to be no distinct correlation between the line FWHM presented in Table \ref{tab1} and the integrated intensity ratios shown in Table \ref{tab2}.
Therefore, the variance of the line ratios among the clumps is mainly attributed to their chemical variety.

\subsection{The localization of hot molecular gases and chemically rich environment}

\label{subsec:hot_environment}
Some of the eight star-forming clumps, especially Clump 1, exhibits chemically rich spectra.
Molecular line emission as a whole is enhanced in the clumps, and we detect enhanced lines of HNCO, COMs, and vibrationally excited HNC.
They also contain rarer and complex species such as H$_2$CS, CH$_3$CCH, and CH$_3$COOH.
The survival of HNCO and COMs suggests not only that such clumps harbor some shielded regions, but also that they are dense molecular gas regions in the early evolutionary phase of star-formation, just after such molecules are produced on dust grains and sublimed into the gas-phase.

\begin{figure}[t!]
\figurenum{5}
\plotone{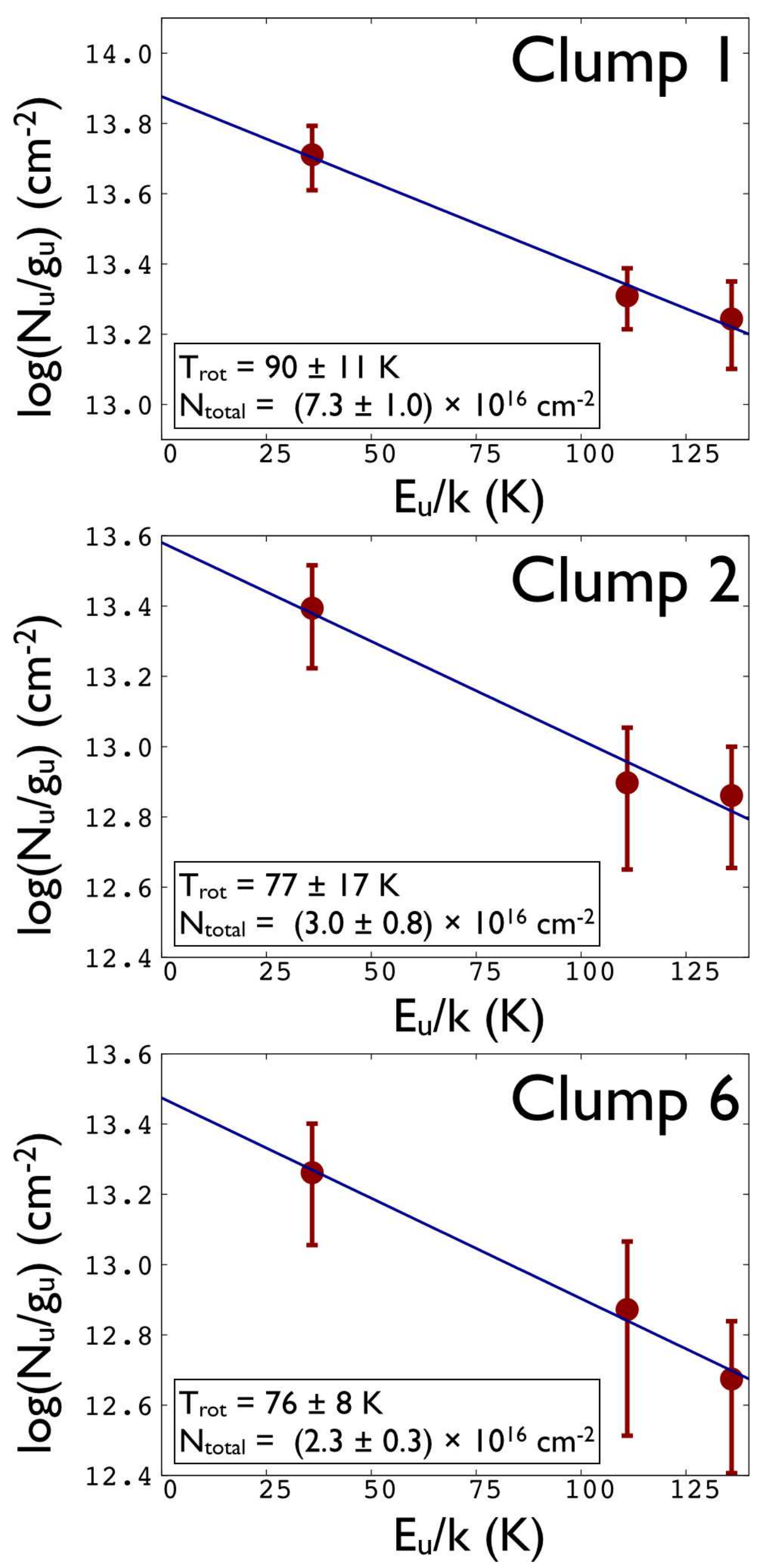}
\caption{Rotation diagram of SO$_2$ for Clumps 1, 2, and 6. 
\label{fig:Trot}}
\end{figure}

For the purpose of investigating the origin of such chemically rich spectra, we perform the rotation diagram analysis \citep{Blake+87} for SO$_2$ to estimate molecular gas temperature.
SO$_2$ is utilized as a tracer of warm and dense molecular gas around high-mass YSOs \citep{Beuther+09}.
We detect three transitions of SO$_2$ in Clumps 1, 2, and 6, where SO$_2$ lines are sufficiently bright for the analysis and comparably various molecules are detected.
All of the three transitions have similar velocity profiles and source sizes.
The rotation diagram is shown in Figure \ref{fig:Trot}, suggesting that the rotation temperature $T_{\mathrm{rot}}$(SO$_2$) $= 90 \pm 11$ K and the column density $N_\mathrm{total}$(SO$_2$) $= (7.3 \pm 1.0) \times 10^{16}$ cm$^{-2}$ for Clump 1.
$T_{\mathrm{rot}}$(SO$_2$) in Clumps 2 and 6 are also derived to be similarly high ($\sim 80$ K).

Our results are consistent with those of previous studies below in that the existence of the hot molecular gas in the heart of NGC 253 is suggested.
Previous observations of NH$_3$ gas in NGC 253 have demonstrated the high temperature environments in its central region.
\cite{Mauersberger+03} have conducted single dish observations with the spatial resolution of $40''$  (corresponding to 680 pc), and have shown that two velocity components have the rotation temperatures of 142 K and 100 K, respectively.
Interferometric observations with the resolution of $6''$ (corresponding to 100 pc) have suggested that the molecular gas within the central kpc of NGC 253 could be best described by a two kinetic temperature model with 130 K and 57 K component \citep{Gorski+17}.
In addition to supporting these previous studies, we also reveal that hot molecular gases ($\sim 100$ K) are localized in a 10-pc scale star-forming clump, and this is strongly suggesting that high molecular temperature plays a major role in bringing the chemically rich environment.
The connection between the hot environment and the chemical richness and their localization within a 10-pc scale clump cannot be disclosed without several-parsec-resolution observations such as our ALMA ones.

Clump 1 is having the brightest and the most chemically rich spectra among the eight star-forming clumps.
Here, in order to interpret the environment of Clump 1 described above, we propose a hypothesis that Clump 1 is a giant hot core cluster, which is a several-parsec-scale aggregate of hot molecular cores \citep{KJ01} where high-mass star-formation takes place.
Hot molecular cores are defined with the following characteristics: small source size ($\leq 0.1$ pc), high density ($\geq 10^6$ cm$^{-3}$), and warm gas and dust temperatures ($\geq 100$ K) (\citealt{Kurtz+00}; \citealt{van_der_Tak+03}; \citealt{Shimonishi+16}).
Owing to the lack of angular resolution of conventional radio telescopes, the observational studies of hot molecular cores were limited to Galactic sources, until \cite{Shimonishi+16} detected the first extragalactic hot core in the LMC with ALMA.
The results of our analysis, which suggest the existence of the hot molecular gases and the chemically rich environment in Clump 1, are consistent with the hypothesis.
Typical hot molecular cores exhibit somewhat higher temperature environments: 
$T_{\mathrm{rot}}$(SO$_2$) is 124 K in Orion KL \citep{Schilke+97}, 184 K in W3(H$_2$O) \citep{HvD97}, and 190 K in ST11 in the LMC \citep{Shimonishi+16}.
However, these hot gas components are distributed just over typical hot core sizes ($\leq 0.1$ pc).
In contrast, Clump 1 in NGC 253 harbors hot molecular gas ($\sim 100$ K) spreading over 10-pc scale.
This result is remarkable because it suggests that a hot environment entirely dominates Clump 1, and it is consistent with the picture that Clump 1 is a giant and rich cluster of hot molecular cores.
Clumps 2 and 6, which are comparably hot to Clump 1 and in the most chemically-rich environments after it, could also be interpreted as giant hot core clusters.

Note that, however, we cannot resolve molecular cores surrounding single protostars in NGC 253, nor estimate the density and dust temperature of Clump 1 even with ALMA;
it is just a hypothesis and we do not exclude other possibilities such as that Clump 1 is a large photodissociation region (PDR). 
However, it might be less likely that Clump 1 is a PDR, since SO$_2$ tend to be easily dissociated in a PDR (e.g. \citealt{Fuente+03}; \citealt{Ginard+12}).
Further observations will make it clear which picture can explain what takes place in Clump 1.
In particular, wider frequency observations of the heart of NGC 253 with comparably high angular resolution will enable us to determine the dust temperature therein and to make robust identifications of the lines of COMs.

This work demonstrates the prominent capability of ALMA;
it can resolve 10-pc scale star-forming clumps in external galaxies and bring their chemical diversity to light.
Future ALMA observations of nearby starburst galaxies with high spatial resolution will lead us to get a closer look into the chemical diversity of starburst regions and to obtain a further understanding of which formation processes and evolutionary stages such regions trace in galaxies.

\acknowledgments
We thank Dr. J. Mangum for his kind advice on the interpretation of our data.
We also would like to express the deepest appreciation to an anonymous referee for giving a number of insightful comments that significantly improved the manuscript.
This paper makes use of the following ALMA data: ADS/JAO.ALMA \#2013.1.00099.S and 2013.1.00735.S. ALMA is a partnership of ESO (representing its member states), NSF (USA) and NINS (Japan), together with NRC (Canada), NSC, ASIAA (Taiwan) and KASI (Republic of Korea), in cooperation with the Republic of Chile. The Joint ALMA Observatory is operated by ESO, NAOJ and NRAO.
RA was supported by the ALMA Japan Research Grant of NAOJ Chile Observatory, NAOJ-ALMA-0138. 
KN and KK acknowledge support from JSPS KAKENHI Grant Numbers 15K05035 and 25247019, respectively. 
This work is supported by NAOJ ALMA Scientific Research Grant Number 2017-06B.

\vspace{5mm}

\facilities{ALMA (NAOJ, NRAO, ESO)}

\software{CASA}

\end{document}